# Buffer gas cooling and trapping of atoms with small magnetic moments

J.G.E. Harris,[1,2] R.A. Michniak,[1,2] S.V. Nguyen,[1,2] W. Ketterle,[1,3] J.M. Doyle[1,2]

[1] *Harvard/MIT Center for Ultracold Atoms, Cambridge Massachusetts 02138*

[2] *Department of Physics, Harvard University, Cambridge Massachusetts 02138*

[3] *Department of Physics, MIT, Cambridge Massachusetts 02139*

Buffer gas cooling was extended to trap atoms with small magnetic moment $\mu$. For $\mu \geq 3\mu_B$, $10^{12}$ atoms were buffer gas cooled, trapped, and thermally isolated in ultra high vacuum with roughly unit efficiency. For $\mu < 3\mu_B$, the fraction of atoms remaining after full thermal isolation was limited by two processes: wind from the rapid removal of the buffer gas and desorbing helium films. In our current apparatus we trap atoms with $\mu \geq 1.1\mu_B$, and thermally isolate atoms with $\mu \geq 2\mu_B$. Extrapolation of our results combined with simulations of the loss processes indicate that it is possible to trap and evaporatively cool $\mu = 1\mu_B$ atoms using buffer gas cooling.

Quantum degenerate atomic gases are produced in a two-stage process. First the atoms are cooled to a temperature low enough for trapping. Second, evaporation cools the atoms further and increases their phase space density. Laser cooling is the method most commonly used in the first stage, but has severe limitations. Laser cooling requires that the atoms have optical cycling transitions at accessible wavelengths. Cooling mixtures of different atoms requires a separate laser system for each species. In addition, the size of laser cooled samples is limited by collisions between excited-state atoms, multiple photon scattering, and technical limits on laser power to a maximum of $\sim 10^{10}$ atoms.[1] As a result, both the range of atomic species and the size of samples which can be laser cooled are limited. Atomic hydrogen has been cooled via thermalization with superfluid helium-coated walls and trapped in much larger numbers;[2] however this approach is not applicable to other species.

Overcoming these limitations on atomic species and sample size would open a number of possibilities for ultracold atoms and molecules, including the study of quantum gases with interactions dominated by anisotropic collisional properties or strong long-range dipolar coupling.[3] Other applications include the trapping of arbitrary mixtures of different species or isotopes, and the study of species particularly well-suited to high precision measurements. Larger samples of cold atoms would be simultaneously in the hydrodynamic and small gas-parameter regimes[4] and could also improve the signal in a wide range of experiments.

Buffer gas cooling was proposed[5] and demonstrated[6] as a way to cool large numbers of atoms of a wide range of species and load them into a magnetic trap. Buffer gas cooling works by thermalizing hot atoms with cold (<1 K) helium gas. If the helium temperature $T_{He}$ and the magnetic trap depth $B_{trap}$ satisfy

$$\mu B_{\text{trap}} > k_B T_{\text{He}} \qquad\qquad (1)$$

where $\mu$ is the atom's magnetic moment, then the thermalized atoms will be trapped. $T_{He}$ must be greater than ~250 mK (~600 mK) to ensure that the $^{3}$He ($^{4}$He) saturated vapor density $P_0$ is adequate (>$10^{15}$ cm$^{-3}$) to thermalize the hot atoms before they reach the cell walls. The deepest traps[7] achieve $B_{\text{trap}} \approx 4$ T. Given these constraints, Eq. (1) can be satisfied for any atom with $\mu \geq 1\mu_B$, i.e., the majority of atomic species.

Once the atoms are trapped, further cooling by evaporation requires that thermal contact between the atoms and the cell walls be broken. This means that the buffer gas must be removed and good vacuum achieved in the cell while the atoms remain trapped. Thus, one must guarantee that the lifetime of the trapped atoms in the presence of the buffer gas is longer than the time required to remove the buffer gas. This requirement is stronger than Eq.(1). Simulations[8] (and the experiments described here) show that the buffer gas limits the trap lifetime to ~ 40 ms for $\mu = 1\mu_B$, increasing to ~ 20 s for $6\mu_B$ atoms (assuming $B_{\text{trap}} = 4$ T and $T_{\text{He}} = 500$ mK). In earlier work $^{4}$He buffer gas was removed by using a dilution refrigerator to cool the cell walls below 200 mK, at which point $P_0$ is sufficiently low that excellent vacuum is achieved. Using this approach ~$10^{12}$ atoms of Eu ($7\mu_B$), Cr ($6\mu_B$), and Mo ($7\mu_B$) were magnetically trapped and thermally isolated.[6,9] In Ref.[10], $10^{8}$ molecules of CaH ($1\mu_B$) were magnetically trapped but not thermally isolated; cooling the cell (and hence isolating the trapped atoms) took several seconds, precluding the use of this approach to thermally isolate species with small $\mu$.

In this paper we demonstrate that atoms with magnetic moments down to 2 $\mu_B$ can be buffer gas trapped and thermally isolated, that thermal isolation can be achieved at cell temperatures of 400 mK and that there are no fundamental impediments to extending this technique to 1 $\mu_B$ atoms. Such an extension would be extremely important as it would greatly increase the number of species which could be trapped in thermal isolation using buffer gas cooling, and would include the species known to have favorable collisional properties for reaching quantum degeneracy. In this work we use a cryogenic valve and *in situ* cryopump to achieve vacuum adequate for evaporative cooling, allowing us to forego the use of a dilution refrigerator.

Three obstacles make it difficult to rapidly achieve good vacuum in a buffer gas cell: limited pumping speed for the $^3$He buffer gas, virtual leaks (*e.g.* from fill lines), and desorption of $^3$He from cell walls. The first two issues are directly addressed in the design of our cell (Fig. 1). A pneumatically actuated, large-aperture cryogenic valve connects the trapping region to a pumping region with ~30 g of activated charcoal cooled to 1.5 K by the $^4$He pot of a $^3$He refrigerator. The estimated conductance of this valve and the volume of the cell ($V = 0.5$ l) give a pumpout time $\tau_{pump} = 50$ ms (100 ms) for $^3$He gas at 500 mK in the viscous (molecular) regime.[11]

$\tau_{pump}$ would be substantially increased if the buffer gas also had to be removed from narrow fill lines connecting the cell to the room-temperature gas handling system. This is avoided by introducing the $^3$He into the trapping region through a high-impedance orifice ($2 \times 10^{-4}$ l/s) from an "antechamber" (Fig. 1) containing a small sorb. Once the trapping region is filled with $^3$He we evacuate both the antechamber and its fill line by cooling this sorb below 2 K. Then we produce the atoms (via laser ablation of a solid target[12]) and after some delay open the valve (which actuates 2 cm in 20 ms), pumping the $^3$He gas onto the large sorb and leaving the magnetic atoms in the trap.

Because the trap lifetime depends upon $\mu$ only via the product $\mu B_{trap}$, we simulate atoms of different effective magnetic moment $\mu_{eff}$ by using Cr ($\mu = 6\mu_B$) and varying $B_{trap}$ (*i.e.*, $\mu_{eff} =$ $\mu B_{\mathrm{trap}} / B_{\mathrm{trap}}^{\max}$ ).[13] The atoms are probed via the absorption of a laser beam tuned to the $^7S_3 - {}^7P_4$ transition (425 nm).

Fig. 2(a) shows the time dependence of the peak density of Cr atoms $n_{\mathrm{Cr}}(t)$ in the trap for $\mu_{\mathrm{eff}} = 3\mu_B$. Similar behavior is seen for larger $\mu_{\mathrm{eff}}$. Initially, the cell is filled to $n_{\mathrm{He}} = 6.2 \times 10^{15}$ cm$^{-3}$.[14] The ablation laser fires at $t = 0$ s, producing hot Cr atoms, $N_{\mathrm{Cr}} = 1.2 \times 10^{12}$ of which thermalize in the maximally trapped $m_J = 3$ state at a temperature $T_{\mathrm{Cr}} = 600$ mK. $T_{\mathrm{Cr}}$ and $n_{\mathrm{Cr}}$ are measured by fitting the trapped atoms' Zeeman-boadened absorption spectrum to that of a Boltzmann distribution of atoms in our magnetic trap, as described in Refs[9,12]. The inset of Figure 2(a) shows one such spectrum and fit.

After the ablation and before the valve opens (0 s < $t$ < 2.5 s in Fig. 2(a)) the Cr atoms are confined by the trap and move diffusively in the $^3$He. Cr is lost from the trap via both dipolar relaxation and evaporation, and $T_{\mathrm{Cr}}$ tracks $T_{\mathrm{cell}}$, the cell temperature (shown in the inset of Figure 3(a)), indicating the trapped Cr is still in thermal contact with the cell walls. At $t = 2.5$ s the valve opens, the $^3$He rushes out of the cell to the large sorb and the trapped Cr is left behind. A small fraction, $x_{\mathrm{wind}}$, of the Cr atoms (too small to be visible in Fig.2(a)) is carried out of the trap with the $^3$He "wind". As one would expect, and in good qualitative agreement with numerical simulations, $x_{\mathrm{wind}}$ increases for increasing $n_{\mathrm{He}}$ and decreasing $B_{\mathrm{trap}}$. This "wind loss" occurs in the ~200 ms immediately after the valve is opened, while most of the $^3$He flows out of the cell.

After the $^3$He is removed, $T_{\mathrm{Cr}}$ remains constant (600 mK) and $n_{\mathrm{Cr}}$ gradually decreases due to 2-body collisions. This can be seen from the fit in Fig. 2(a) of the form $n_{\mathrm{Cr}}(t) = n_0/(1 + n_0 \Gamma_2 t/8)$. The two fitting parameters are $n_0 = n_{\mathrm{Cr}}(0)$ and $\Gamma_2$, the two-body rate coefficient. The value of $\Gamma_2$ agrees well with previous measurements. The trap loss is due to dipolar relaxation to untrapped $m_J$ states and causes heating of the cloud. Cooling via evaporation over the top of the trap balances this heating and leads to an equilibrium in which $\eta = \mu B_{\text{trap}}/k_B T_{\text{Cr}}$ is independent of time.[15]

We lower $T_{\text{Cr}}$ by decreasing $B_{\text{trap}}$ (cooling occurs via evaporation and adiabatic expansion), as shown in Fig. 2(b). Here $10^{12}$ Cr atoms were trapped as described above, and after opening the valve $B_{\text{trap}}$ was reduced by a factor of 50 to 39 mT. The inset of Fig. 2(b) shows a spectrum taken after this cooling with a fit giving $T_{\text{Cr}} = 47$ mK. The decay of $n_{\text{Cr}}$ (squares in Fig. 2(b)) is exponential and hence no longer dominated by two-body processes, but rather by one-body. Fitting this data to $n_{Cr}(t) = n_{Cr}(0)e^{-t/\tau_1}$ gives a one-body time constant $\tau_1 = 37$ s. One-body loss is absent in Fig. 2(a) because there $B_{\text{trap}} = 1.95$ T and so the Cr see a trap substantially deeper than the mean thermal energy of a background gas atom $k_B T_{\text{cell}}$. For $T_{\text{cell}} = 500$ mK only one Cr-$^3$He collision in $10^{28}$ ejects a Cr atom from the trap. However for $B_{trap} = 39$ mT, this ratio is roughly unity and $\tau_1$ provides a direct measure of the background gas density $n_{\text{He}} = (\tau_1 \sigma v)^{-1} = 4.5 \times 10^8$ cm$^{-3}$ where $\sigma = 10^{-14}$ cm$^2$ is the Cr-$^3$He elastic cross section and $v = 5300$ cm/s is the mean thermal velocity of the $^3$He. This level of background gas does not impede the evaporative cooling (note that a 30-second lifetime is adequate to achieve BEC in other experiments[16]).

The fact that $n_{\text{He}}$ remains roughly constant after reaching $4.5 \times 10^8$ cm$^{-3}$ implies an influx of $2.3 \times 10^{12}$ $^3$He atoms/s into the cell volume. This influx is due to desorption of the thin $^3$He film which coats the cell each time it is filled with buffer gas. We demonstrate that this film is responsible for the 37 second lifetime described above by "baking out" the cell (in analogy with the bake-out of water films in room-temperature UHV systems). After filling the cell with buffer gas, ablating, trapping $10^{12}$ Cr atoms, and removing the buffer gas all with $B_{\text{trap}} = 1.95$ T we heat the cell from 520 mK to 650 mK for 30 s. We then allow the cell to cool back to 470 mK. Only then do we evaporatively cool the Cr atoms by decreasing $B_{\text{trap}}$ to 39 mT. By following this procedure we find that $\tau_1$ is increased to ~$10^3$ s (circles in Fig. 2(b)).

The behavior of the trapped atoms depends strongly upon $\mu_{\text{eff}}$. For $\mu_{\text{eff}} < 3\mu_B$ we observe enhanced atom loss in the first seconds after the valve opens. Figure 3(a) shows $n_{\text{Cr}}(t)$ for $\mu_{\text{eff}} = 2.45\mu_B$. Ablation occurs at $t = -1.2$ s and the valve opens at $t = 0$ s. There is ~75% atom loss in the first 5 s after the valve opens. After this $n_{\text{Cr}}$ remains nearly constant, decaying only very slowly via Cr-Cr collisions as in Fig. 2(a). The loss seen in Fig. 3(a) can not be attributed to the $^3$He wind, which occurs only in the first 200 ms after the valve opens. Instead, this loss is due to collisions of Cr with $^3$He atoms desorbed from the cell wall.

We model this desorption using simple gas dynamics: the $^3$He desorption rate $\dot{n}_d = P(d)fA/V\sqrt{2\pi k_B T_{\text{cell}}(t)m}$ where $m$ is the $^3$He mass, $f$ is the sticking fraction (taken to be 0.75),[17] $A = 420$ cm$^2$ is the surface area of the cell, and the vapor pressure $P$ is related to the film thickness $d$ and $P_0$ by the FHH expression[18] $P = P_0\exp(-\alpha/k_B T_{\text{cell}}(t)d^3)$. We assume the van der Waals coefficient[19] $\alpha = 1900$ KÅ$^3$ and use the measured $T_{\text{cell}}(t)$ shown in the inset of Fig. 3(a). Because of this desorption the film thins at a rate

$$\dot{d} = \frac{d_0}{N_0}\left(-\dot{n}_d V + n_{He}Af\sqrt{\frac{k_B T_{\text{cell}}}{2\pi m}}\right)$$

where the first term in the parentheses is the rate of desorption from the film and the second term is the rate of adsorption from the $^3$He vapor. $N_0$ is the number of $^3$He atoms per monolayer and $d_0$ is the thickness of a monolayer. Finally, the rate of change of the $^3$He density in the cell is the difference of the outflux of atoms (to the cryopump) and the influx of atoms (off the cell walls):

$\dot{n}_{\text{He}} = -\dfrac{n_{\text{He}}}{\tau_{\text{pump}}} + \dot{n}_d$ . We solve this series of equations numerically for $d(t)$, $\dot{n}_d(t)$ and $n_{\text{He}}(t)$.

The results of this model are shown in Fig.3(b), and account qualitatively for the data. At first $n_{He}$ rapidly drops three orders of magnitude (as $e^{-t/\tau_{pump}}$) but then slows due to the desorbing film.[20] This slowing prolongs the thermal contact between the Cr and the cell, and for small $\mu_{eff}$ gives rise to the loss seen in Fig. 3(a). The shut-off of this loss after 5 s is consistent with the simulations in Ref.[8] which show that loss due to thermal contact with the cell walls decreases rapidly for $n_{He} < 10^{13}$ cm$^{-3}$.

The value of $n_{He}$ calculated at long times ($10^8$ cm$^{-3}$, inset of Fig. 3(b)) is consistent with the lifetime $\tau_1$ observed in the no-bake-out evaporative cooling data of Fig. 2(b). Including the "bake-out" routine in the simulation decreases the long-time value of $n_{He}$, consistent with Fig. 2(b).

The effect of the film depends strongly upon $T_{cell}(t)$: lower $T_{cell}$ means less energetic $^3$He-Cr collisions, and a drop in $T_{cell}$ after the valve opens cryopumps the film (lowering $n_{He}$). Calculations of $n_{He}(t)$ for slightly improved base temperature and cooling rate of the cell are shown as dashed lines in Figs. 3(a) & (b) and indicate that modest improvements in $T_{cell}(t)$ can result in much faster breaking of thermal contact.

Our results are summarized in Figure 4. The square points show the number of trapped atoms 300 ms after the valve is opened. These atoms have survived the "wind" and the removal of ~99% of the buffer gas (Fig. 3(b)), but are still in thermal contact with the cell walls. The number of atoms remaining in the trap 10 s after the valve opens is shown as circles. These atoms are thermally isolated and so can be cooled further by evaporation. The difference between $N_{Cr}$ at 300 ms and 10 s is appreciable only for $\mu_{eff} < 3\mu_B$ and is overwhelmingly due to the desorbing film. Small improvements in the thermal performance of the cell can lessen the effect of the film (as described above), and can ensure that atoms trapped at 300 ms remain trapped. Further mitigation of the film can be achieved by coating the cell walls with alkali metal (*e.g.*, by *in situ* laser ablation). The small binding energy of He on Na or Rb implies that the entire He film should desorb on a time scale shorter than $\tau_{\text{pump}}$.

For $\mu_{\text{eff}} \geq 3 \; \mu_{\text{B}}$, the number of trapped and thermally isolated atoms is only limited by the ablation yield. Ablation of less refractory metals such as Na produces ~100 times more thermalized atoms.[21] We have also found that for $\mu_{\text{eff}} = 6\mu_B$ it is possible to trap and thermally isolate atoms with $T_{\text{cell}} > 1.5$ K, allowing the study of large-$\mu$ atoms using only a pumped [4]He cryostat.

In conclusion, we have extended buffer gas trapping to new regimes, allowing the cooling and trapping of a much wider range of atoms and molecules than has previously been possible.

We acknowledge the assistance of Nathaniel Brahms, Joel Helton, Andrew Jayich and Bernard Zygelman. This work was supported by the NSF through the Harvard/MIT Center for Ultracold Atoms.

FIG. 1: Schematic cross-section of the buffer gas cell with the valve closed (left) and open (right). *A*: fill line; *B*: small sorb; *C*: antechamber; *D*: high-impedance orifice; *E*: trapping region; *F*: window; *G*: Cr ablation target; *H*: large sorb; *I*: valve shaft; *J*: pumping region; *K*: mirror; *L*: trapped atoms.

FIG. 2: Trapping and evaporative cooling of atoms with $3\mu_B$ effective magnetic moment. (a) Decay of the peak density of trapped Cr. Solid line: fit to two-body loss. Inset: spectrum of trapped Cr 5 s after the valve opens. The fit gives $T_{Cr}$ = 600 mK, $n_{Cr}$ = $1.2 \times 10^{12}$ cm$^{-3}$ and $N_{Cr}$ = $5.5 \times 10^{11}$ atoms. (b) $n_{Cr}(t)$ after evaporative cooling. The bake-out data has been scaled to match the no-bake-out data at $t$ = 0. The lines are fits to exponential decay and show that baking the cell increases the trap lifetime from 37 s to 995 s. Inset: Spectrum of trapped Cr atoms after evaporative cooling. The fit gives $T_{Cr}$ = 47 mK, $n_{Cr}$ = $3.6 \times 10^{9}$ cm$^{-3}$ and $N_{Cr}$ = $9.8 \times 10^{10}$ atoms. Only the peak at –0.6 GHz (due to $^{52}$Cr) is fit. The shoulder at –0.2 GHz is due to $^{53}$Cr.

FIG. 3: Trap loss due to the $^3$He film. (a) Decay of the peak density of trapped Cr for $\mu_{eff}$ = 2.45 $\mu_B$. The inset shows the cell temperature. Solid line: measured $T_{cell}$. The initial rise is due to the ablation. The cell cools to 420 mK after the valve opens. Dashed line: $T_{cell}$ calculated for a base temperature of 320 mK and a cooling rate four times that of the present cell. (b) Solid line: $^3$He density in the cell as a function of time, calculated using the model described in the text and the measured $T_{cell}(t)$. Dashed line: $n_{He}(t)$ calculated for an improved $T_{cell}(t)$ (dashed line in inset of (a)). Dotted line: $n_{He}(t)$

calculated for the experiments of Ref.[9], in which the buffer gas was removed by lowering $T_{cell}$. Inset: the same calculations for longer times.

FIG 4: Trapping as a function of $\mu_{eff}$. Squares: number of atoms trapped 300 ms after the valve opens. Circles: number of atoms trapped and thermally isolated 10 s after the valve opens.

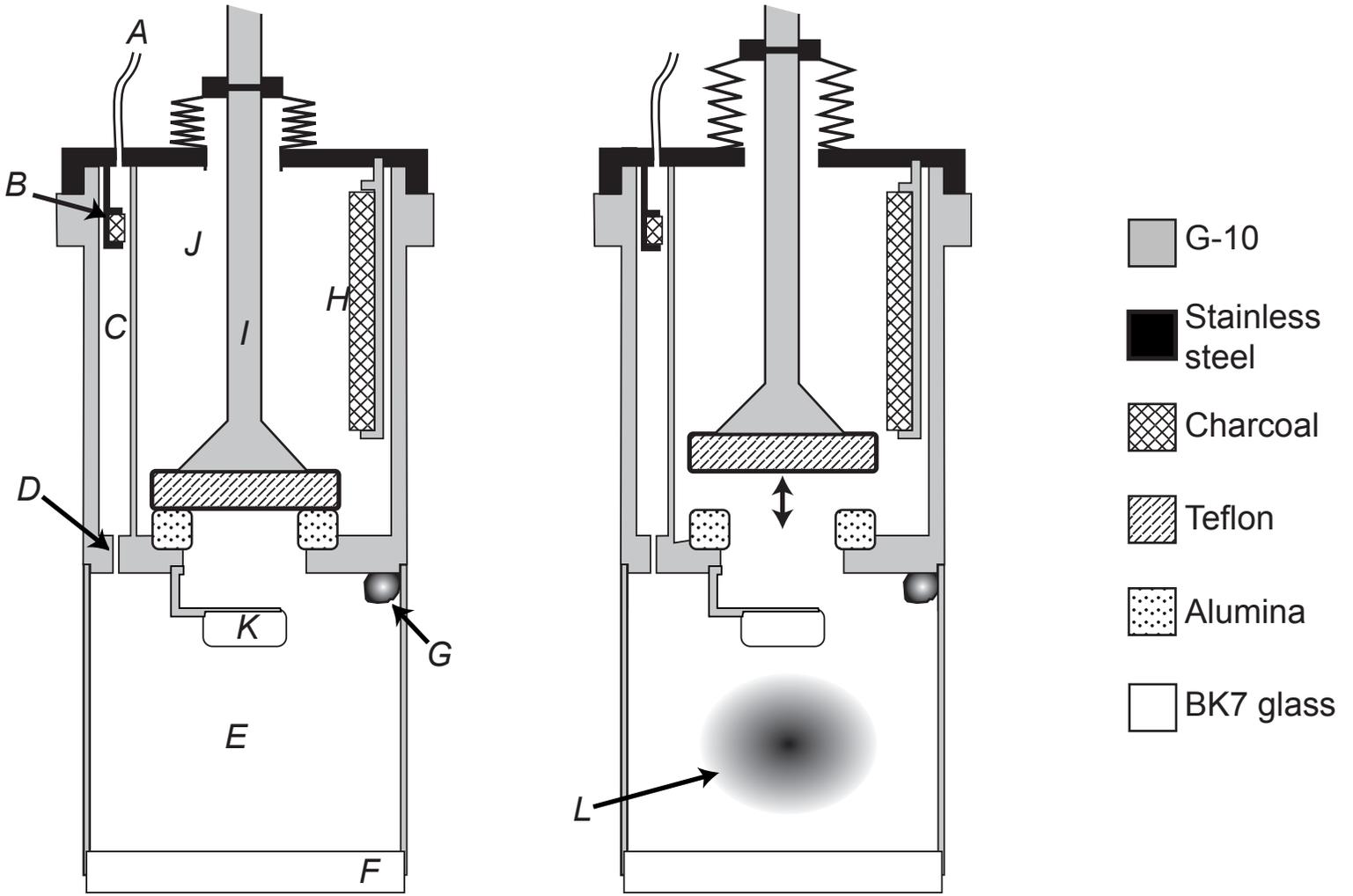

G-10

Stainless
steel

Charcoal

Teflon

Alumina

BK7 glass

J.G.E. Harris *et al*.
Figure 1

**(a)**

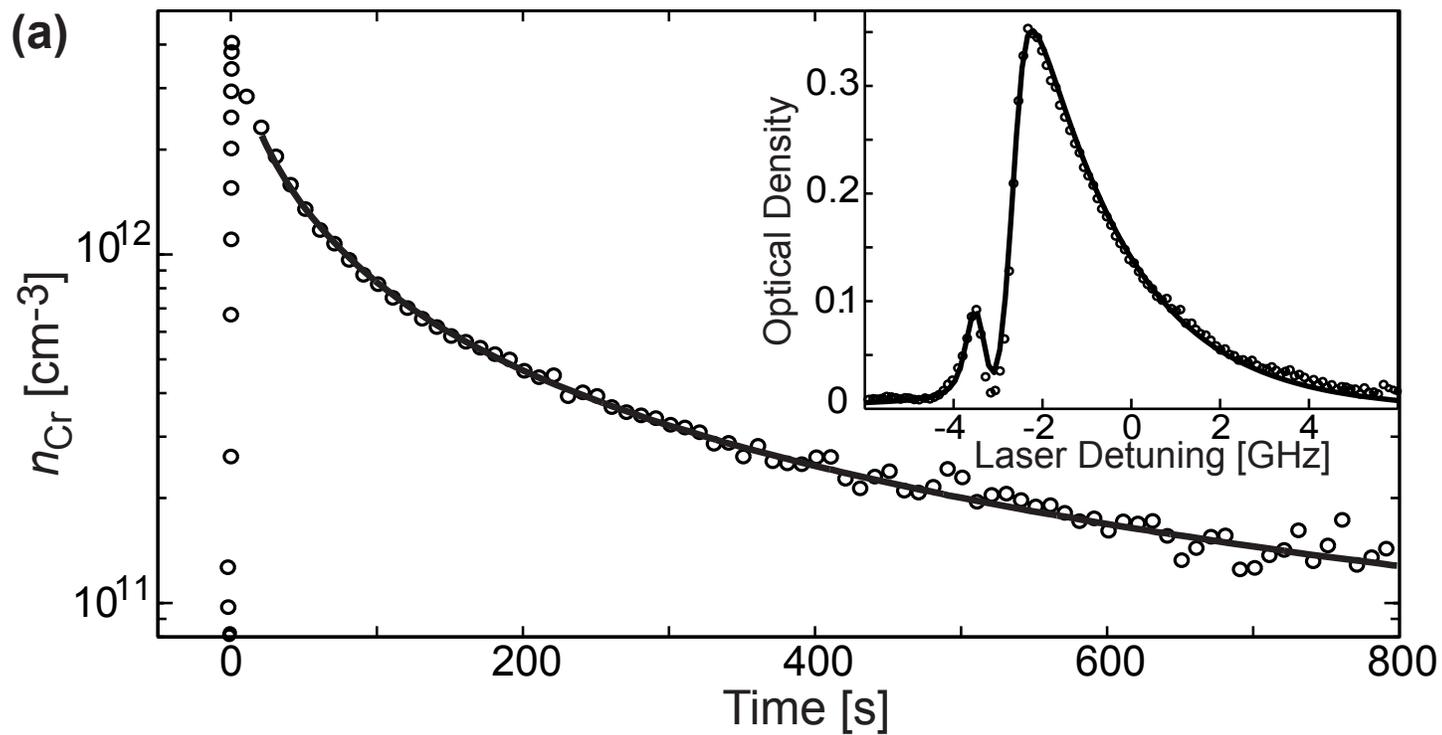

**(b)**

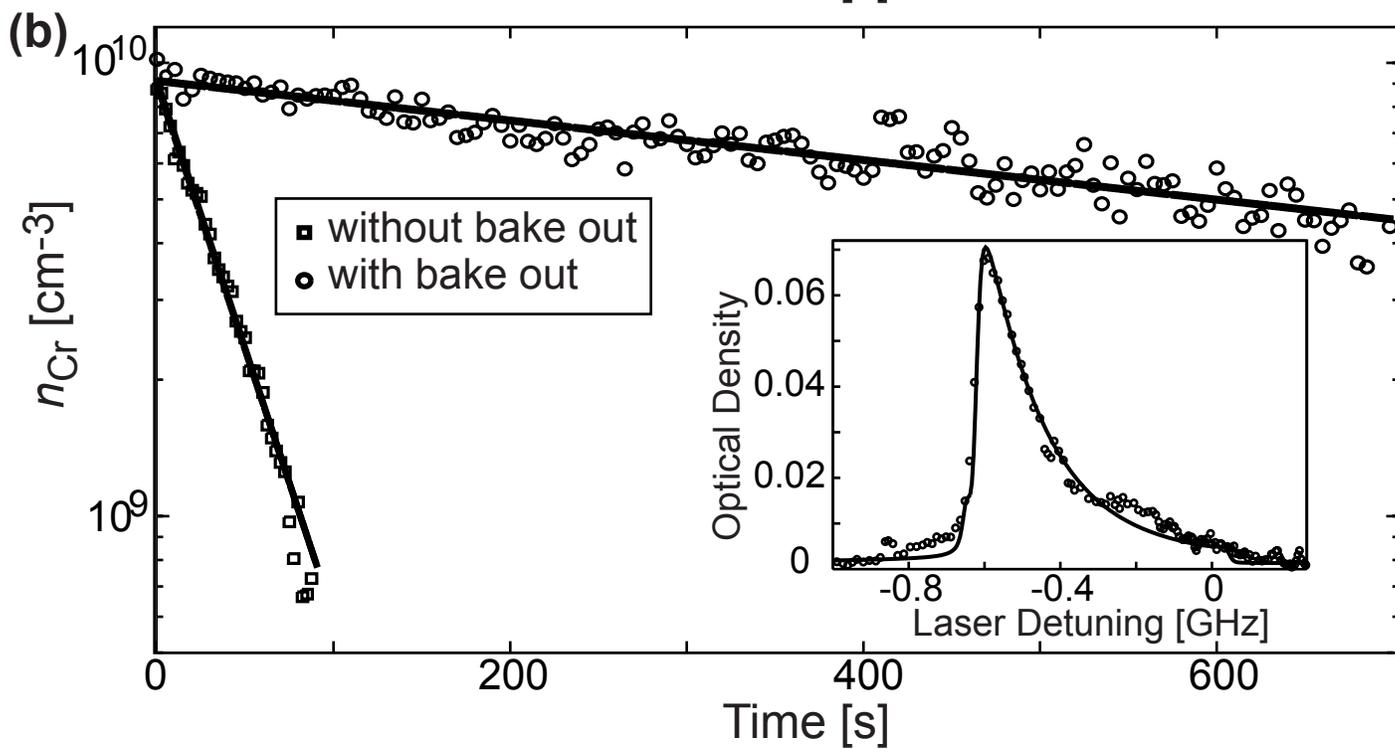

□ without bake out
○ with bake out

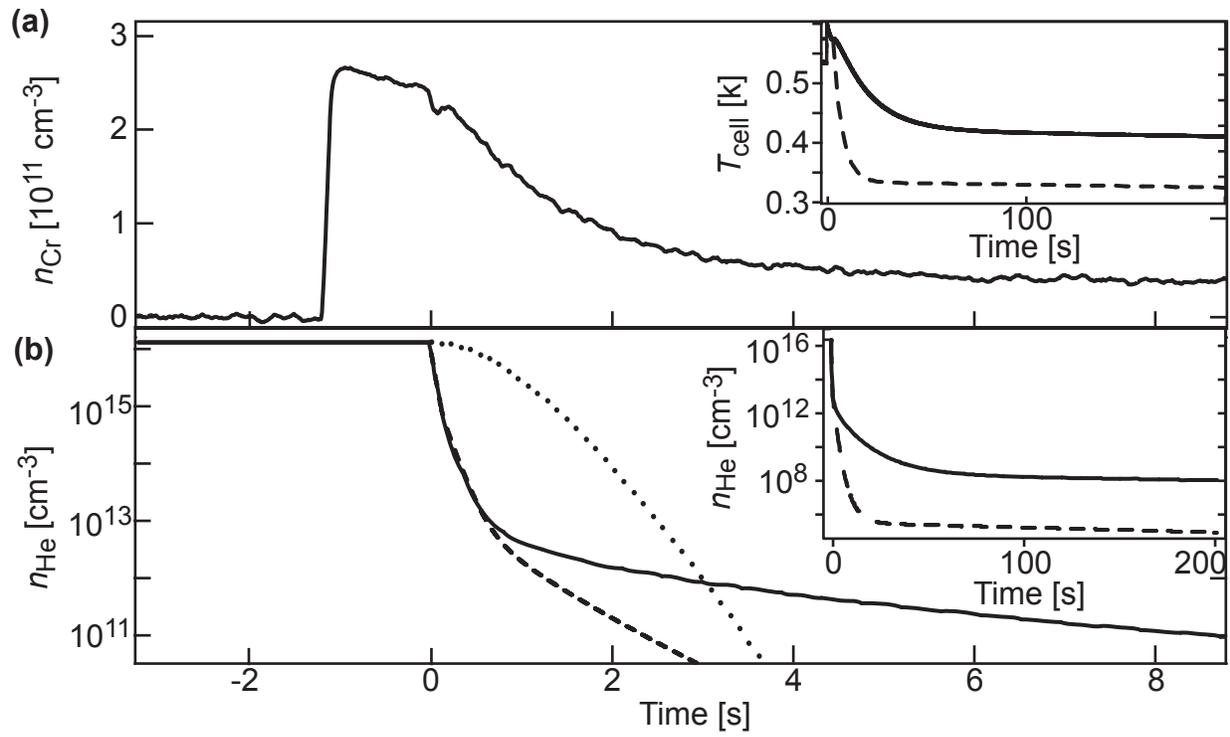

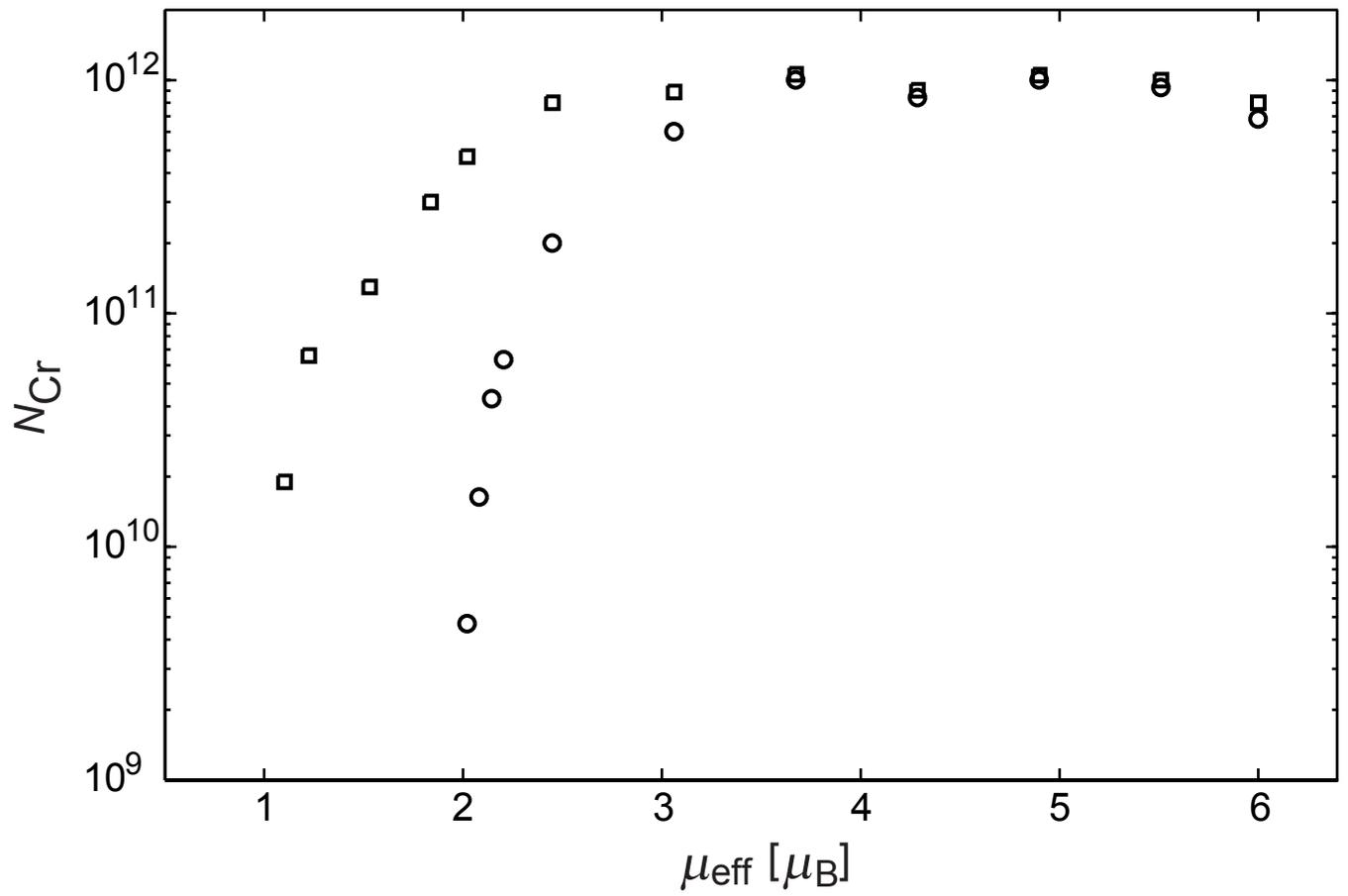

J.G.E. Harris *et al.*
Figure 4